\documentclass[a4paper]{VisionStyle}
\usepackage{epsfig}

\advance\voffset by -1.5cm

\begin{document}

\title{The discovery of X--rays from Venus with Chandra}

\author{K.\,Dennerl\inst{1} \and V.\,Burwitz\inst{1} \and
        J.\,Englhauser\inst{1} \and C.\,Lisse\inst{2} \and S.\,Wolk\inst{3}}

\institute{Max--Planck--Institut f\"ur extraterrestrische Physik,
           Giessenbachstra{\ss}e, D--85748 Garching, Germany
          \and
           University of Maryland,
           Department of Astronomy, College Park, MD 20742
          \and
           Chandra X--Ray Center,
           Harvard--Smithsonian Center for Astrophysics,
           60 Garden Street, Cambridge, MA 02138}

\maketitle 

\begin{abstract}

On January 10 and 13, 2001, \object{Venus} was observed for the first time
with an X--ray astronomy satellite. The observation, performed with the
ACIS--I and LETG\,/\,\mbox{ACIS--S} instruments on Chandra, yielded data of high
spatial, spectral, and temporal resolution. Venus is clearly detected as a
half--lit crescent, with considerable brightening on the sunward limb. The
morphology agrees well with that expected from fluorescent scattering of solar
X--rays in the planetary atmosphere. The radiation is observed at discrete
energies, mainly at the O--K$_\alpha$ energy of 0.53~keV. Fluorescence
radiation is also detected from C--K$_\alpha$ at 0.28 keV and, marginally, from
N--K$_\alpha$ at 0.40~keV. An additional emission line is indicated at
0.29~keV, which might be the signature of the C $1s\to\pi^{\star}$ transition
in CO$_2$ and CO. Evidence for temporal variability of the X--ray flux was
found at the $2.6\,\sigma$ level, with fluctuations by factors of a few times
indicated on time scales of minutes. All these findings are fully consistent
with fluorescent scattering of solar X--rays. No other source of X--ray
emission was detected, in particular none from charge exchange interactions
between highly charged heavy solar wind ions and atmospheric neutrals, the
dominant process for the X--ray emission of comets. This is in agreement with
the sensitivity of the observation.


\end{abstract}

\begin{figure}[!ht]
\vspace*{3mm}
\begin{center}
\includegraphics[clip,width=8.5cm]{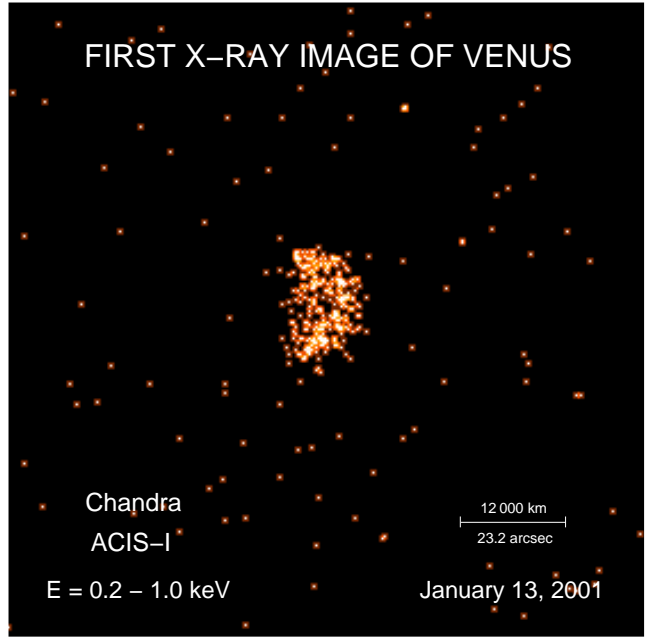}
\end{center}
\caption{Chandra ACIS--I image of Venus.}
\label{kdennerl-B1_fig:xfirst}
\end{figure}

\section{Introduction}

\begin{figure}[!ht]
\begin{center}
\includegraphics[clip,width=8.1cm]{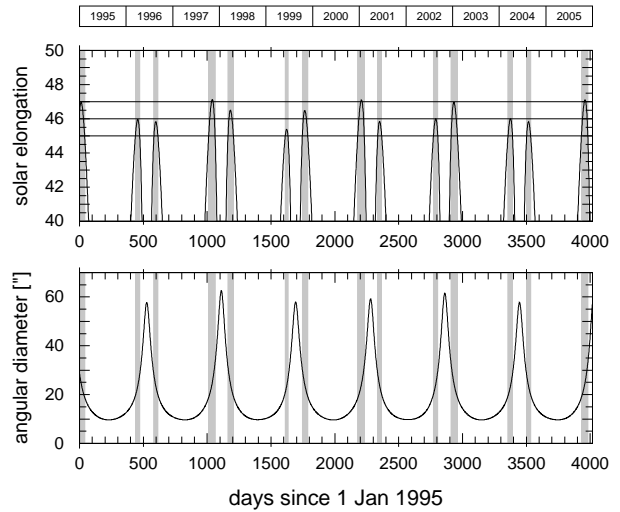}
\end{center}
\vspace*{-3mm}
\caption{Observing constraints for Venus from 1995 to 2005. Shaded areas
highlight the periods when the solar elongation of Venus exceeds $45\degr$,
the minimum angle for a Chandra observation. The observing window in
January 2001 was the most favourable one after the launch of Chandra.}
\label{kdennerl-B1_fig:venus5}
\end{figure}

\begin{figure}[!ht]
\begin{center}
\includegraphics[clip,width=6.8cm]{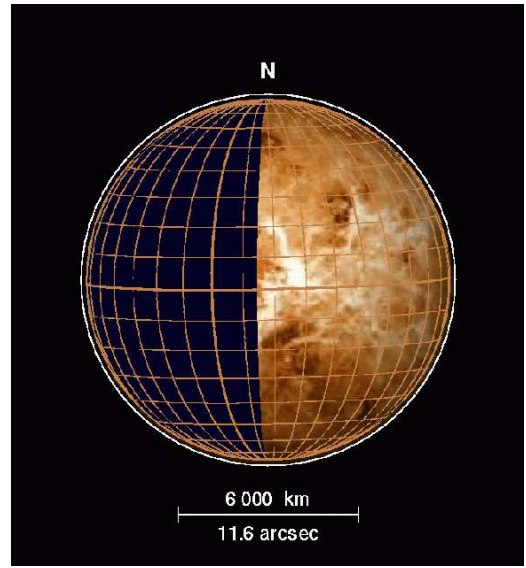}
\end{center}
\caption{Viewing geometry of Venus during the ACIS--I observation on 13 Jan
2001. An equatorial grid $(\ell,b)$, with $\ell=0^{\circ}$ and $b=180^{\circ}$
marked by thick lines, is superimposed on a topographic map obtained by the
Magellan mission. The dark area marks the geometrical shadow, and the outer
white circle the extent of the model atmosphere.}
\label{kdennerl-B1_fig:v7gg}
\end{figure}

In January 2001 we observed \object{Venus} for the first time ever with an
X--ray telescope (Fig.\,\ref{kdennerl-B1_fig:xfirst},
Tab.\,\ref{kdennerl-B1_tab:obscxo}). Orbiting the Sun at heliocentric
distances of 0.718\,--\,0.728 astronomical units (AU), the angular separation
of Venus from the Sun, as seen from Earth, never exceeds 47.8 degrees
(Fig.\,\ref{kdennerl-B1_fig:venus5}). While the observing window of imaging
X--ray astronomy satellites is usually restricted to solar elongations of at
least $60\degr$, Chandra is the first such satellite which is able to observe
as close as $45\degr$ from the limb of the Sun. Thus, with Chandra an
observation of Venus with an imaging X--ray astronomy satellite became
possible for the first time. The observation was scheduled to take place
around the time of greatest eastern elongation, when Venus was $47\degr$ away
from the Sun (Fig.\,\ref{kdennerl-B1_fig:venus5}). At that time it appeared
optically as a very bright (-4.4~mag), approximately half--illuminated
crescent with a diameter of $23\arcsec$ (Tab.\,\ref{kdennerl-B1_tab:obsgeo},
Figs.\,\ref{kdennerl-B1_fig:v7gg},\,\ref{kdennerl-B1_fig:simsum}e).

This observation led to the discovery of X--rays from Venus. It provided not
only the first X--ray image of this Earth--like planet
(Fig.\,\ref{kdennerl-B1_fig:xfirst}), but also spectra and
lightcurves, which together give a consistent picture about the origin of the
X--rays. The main scientific results will appear in
\cite*{kdennerl-B1:den02}. Here we summarize them and present additional
information.

\section{Planning the observation}

Venus is the celestial object with the highest optical surface brightness
after the Sun and a highly challenging target for an X--ray observation with
Chandra, as the X--ray detectors there (CCDs and microchannel plates) are also
sensitive to optical light. Suppression of optical light is achieved by
optical blocking filters which, however, must not attenuate the X--rays
significantly. The observation was originally planned to use the
back--illuminated ACIS--S3 CCD, which has the highest sensitivity to soft
(E$<$1.4~keV) X--rays, for direct imaging of Venus, utilizing the intrinsic
energy resolution for obtaining spectral information. Before the observation
was scheduled, however, it turned out that the optical filter on this CCD
would not be sufficient for blocking the extremely high optical flux from
Venus. Therefore, half of the observation (obsid 583,
cf.\,Tab.\,\ref{kdennerl-B1_tab:obscxo})
was performed with the front--illuminated CCDs of the ACIS--I array (I1 and
I3), which are less sensitive to soft X--rays, but which are also
significantly less affected by optical light contamination.

In order to avoid any ambiguity in the X--ray signal due to residual optical
light, we utilized the low energy transmission grating (LETG) for the other
half of the observation (obsid 2411 and 2414,
cf.\,Tab.\,\ref{kdennerl-B1_tab:obscxo}). The LETG played an an essential role
for the Venus observation. Not only did it allow us to obtain a high
resolution X-ray spectrum, but it provided also an efficient way of
diffracting the extremely intense optical light to areas outside the CCDs, so
that optical photons would not interfere with the X-ray observation. With this
technique we could undoubtedly prove that X-rays were detected from Venus,
despite the fact that the X-ray intensity turned out not to exceed one
ten-billionth of the optical intensity. The combination of direct imaging and
spectroscopy with the LETG made it possible to obtain complementary spatial
and spectral information within the available total observing time of 6.5
hours.

\begin{table}
\caption[]{Journal of observations}
\label{kdennerl-B1_tab:obscxo}
\tabcolsep=0.7mm
\begin{tabular}{ccccc}
\hline
\noalign{\smallskip}
obsid & date & time~~[UT] & net exp & instrument \\
\hline
\noalign{\smallskip}
2411 & 2001\,Jan\,10 & 19:32:47\,--\,21:11:55 & \phantom{0}5\,948~s &
                                                     LETG/ACIS--S \\
2414 & 2001\,Jan\,10 & 21:24:30\,--\,23:00:26 & \phantom{0}5\,756~s &
                                                     LETG/ACIS--S \\
\phantom{0}583 & 2001\,Jan\,13 & 12:39:51\,--\,15:57:40 & 11\,869~s &
                                                        ACIS--I \\
\noalign{\smallskip}
\hline
\end{tabular}
\end{table}

\begin{table}
\caption[]{Observing geometry of Venus}
\label{kdennerl-B1_tab:obsgeo}
\begin{tabular}{cccccccc}
\hline
\noalign{\smallskip}
obsid & $r$ & $\Delta$ & phase & elong & diam & $v_r$ \\
 & [AU] & [AU] & [$^{\circ}$] & [$^{\circ}$] & [$''$] & [km/s] \\
\hline
\noalign{\smallskip}
2411 & 0.722 & 0.734 & 85.0 & 47.0 & 22.7 & -12.8 \\
2414 & 0.722 & 0.734 & 85.0 & 47.0 & 22.8 & -12.8 \\
\phantom{0}583 & 0.721 & 0.714 & 86.5 & 47.1 & 23.4 & -12.8 \\
\noalign{\smallskip}
\hline
\end{tabular}\\[0.8ex]
\vbox{\hsize=0.93\hsize
$r$: distance from Sun, $\Delta$: distance from Earth,
phase: angle Sun--Venus--Earth, elong: angle Sun--Earth--Venus,
diam: angular diameter, $v_r$: radial velocity between Venus and Earth}
\end{table}

\begin{figure}[!ht]
\begin{center}
\includegraphics[clip,width=7.5cm]{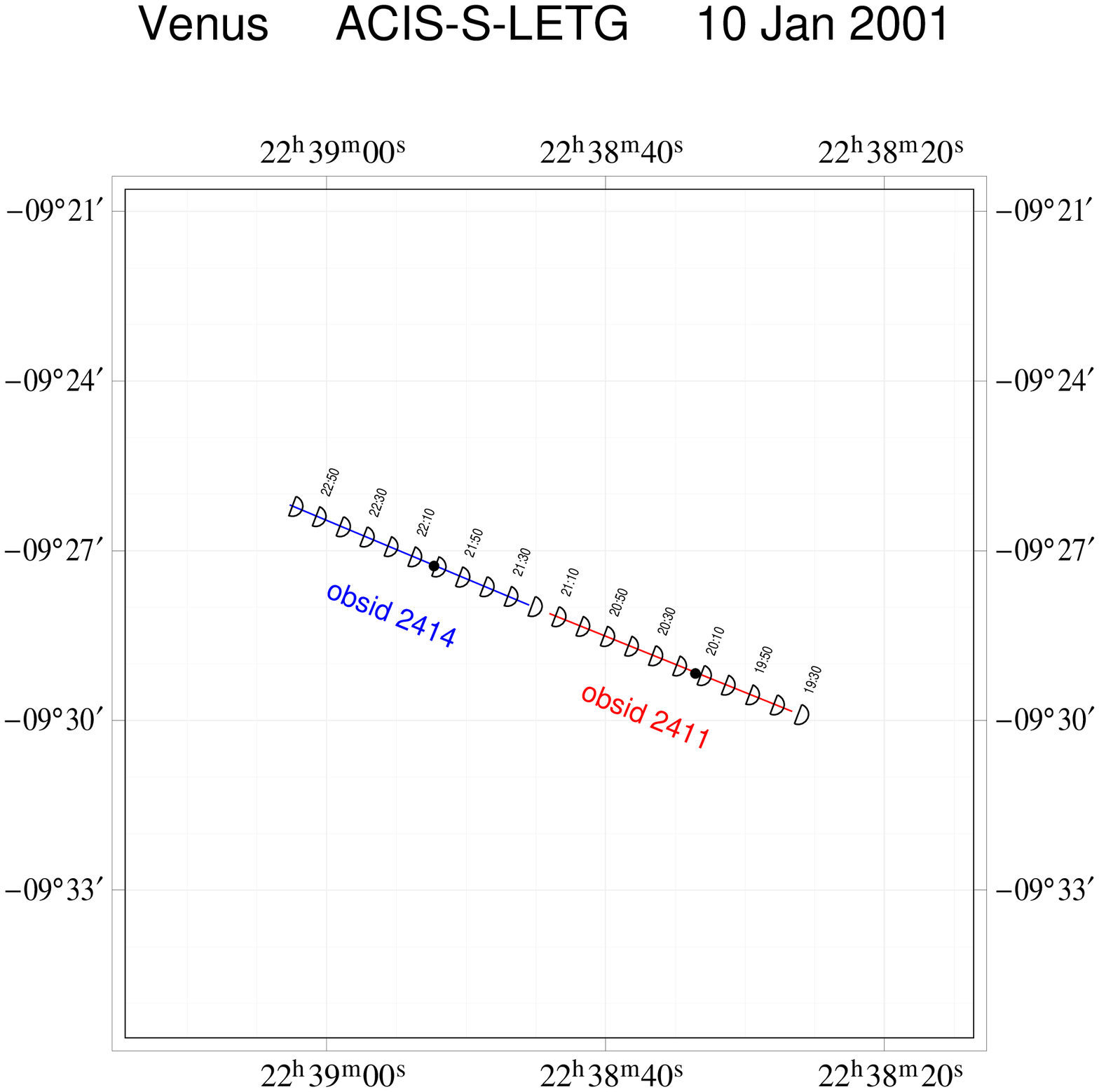}
\end{center}
\caption{Venus during the LETG/ACIS--S
observation on 10 Jan 2001, as seen from
Chandra. The displacement from the geocentric position due to the parallax of
the Chandra orbit varied between $2\farcm1$ and $2\farcm9$. This observation
was split into two parts, to keep Venus in the ACIS--LETG field of view. Black
dots show the two Chandra pointing directions. Images of the Venus crescent are
plotted every 10~minutes. The size and orientation of Venus is drawn to scale.}
\label{kdennerl-B1_fig:acis-s}
\end{figure}

\begin{figure}[!ht]
\begin{center}
\includegraphics[clip,width=7.5cm]{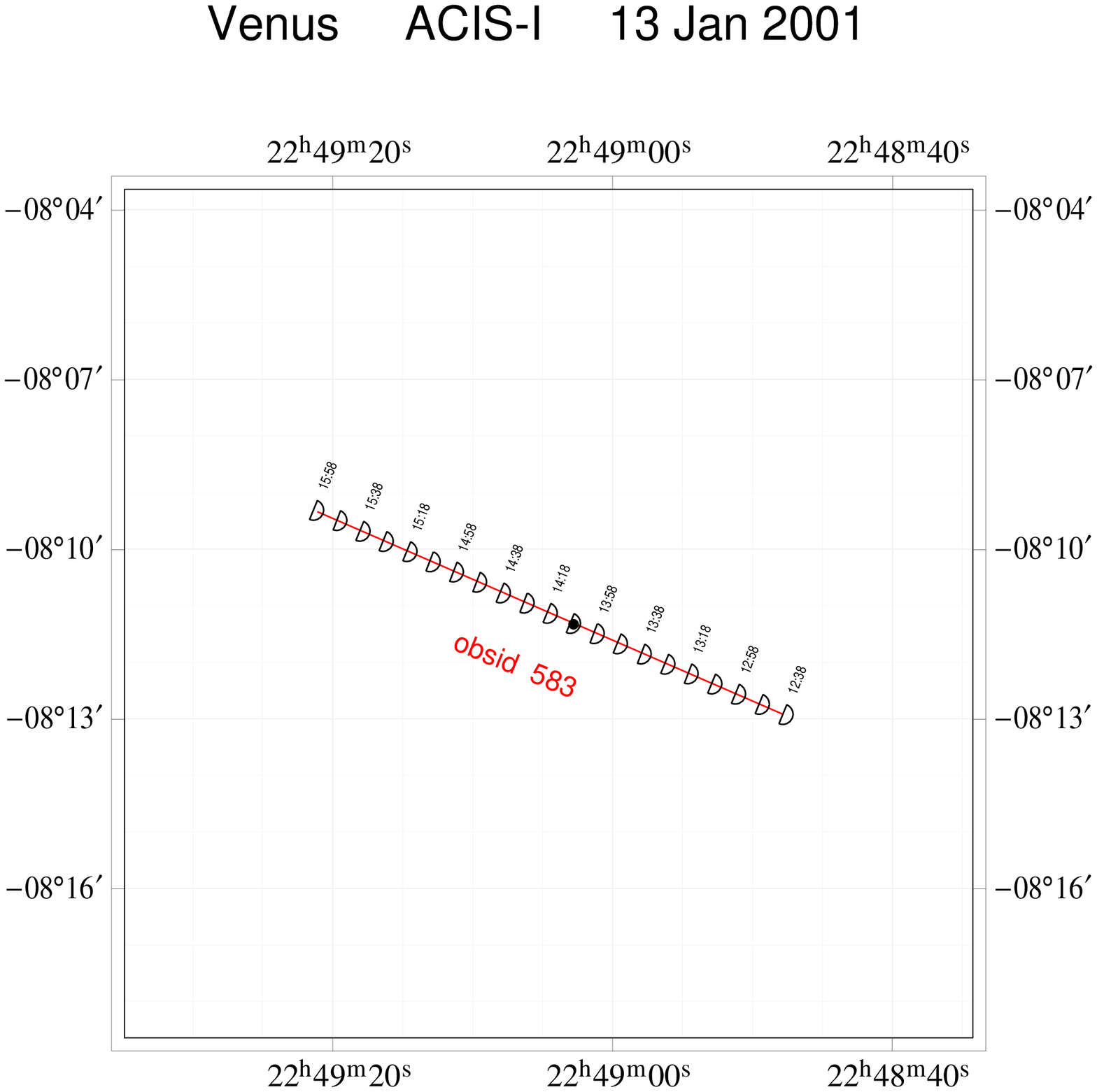}
\end{center}
\caption{Same as Fig.\,\ref{kdennerl-B1_fig:acis-s}, but for the
ACIS--I observation on 13 Jan 2001. Here the parallactic displacement from the
geocentric position varied between $2\farcm6$ and $3\farcm2$.}
\label{kdennerl-B1_fig:acis-i}
\end{figure}

\begin{figure}[!ht]
\begin{center}
\includegraphics[clip,width=7.0cm]{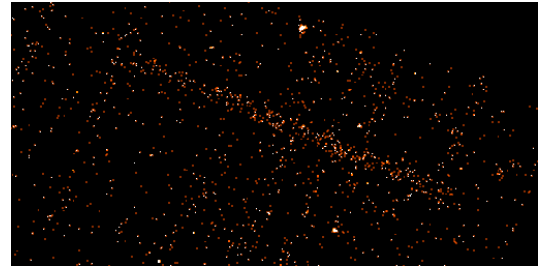}
\end{center}
\caption{Venus moving across ACIS--I during the observation on 13 Jan 2001.
Photons with Chandra standard grades, detected at $E<1\mbox{ keV}$, are
displayed in celestial coordinates.}
\label{kdennerl-B1_fig:acis_i_trail_g}
\end{figure}

\begin{figure}[!ht]
\begin{center}
\includegraphics[clip,width=7.0cm]{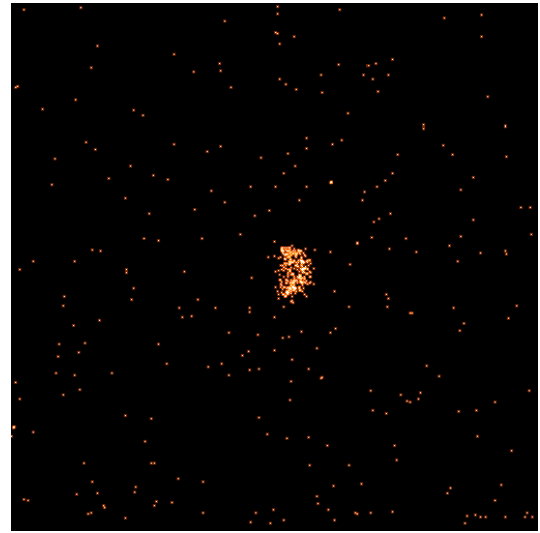}
\end{center}
\caption{ACIS--I image of Venus in the (instrumental) energy range
0.2\,--\,1.0~keV, obtained after reprojecting the photons into the
rest frame of Venus.
}
\label{kdennerl-B1_fig:v00}
\end{figure}

\begin{figure}[!hb]
\begin{center}
\includegraphics[clip,width=7.0cm]{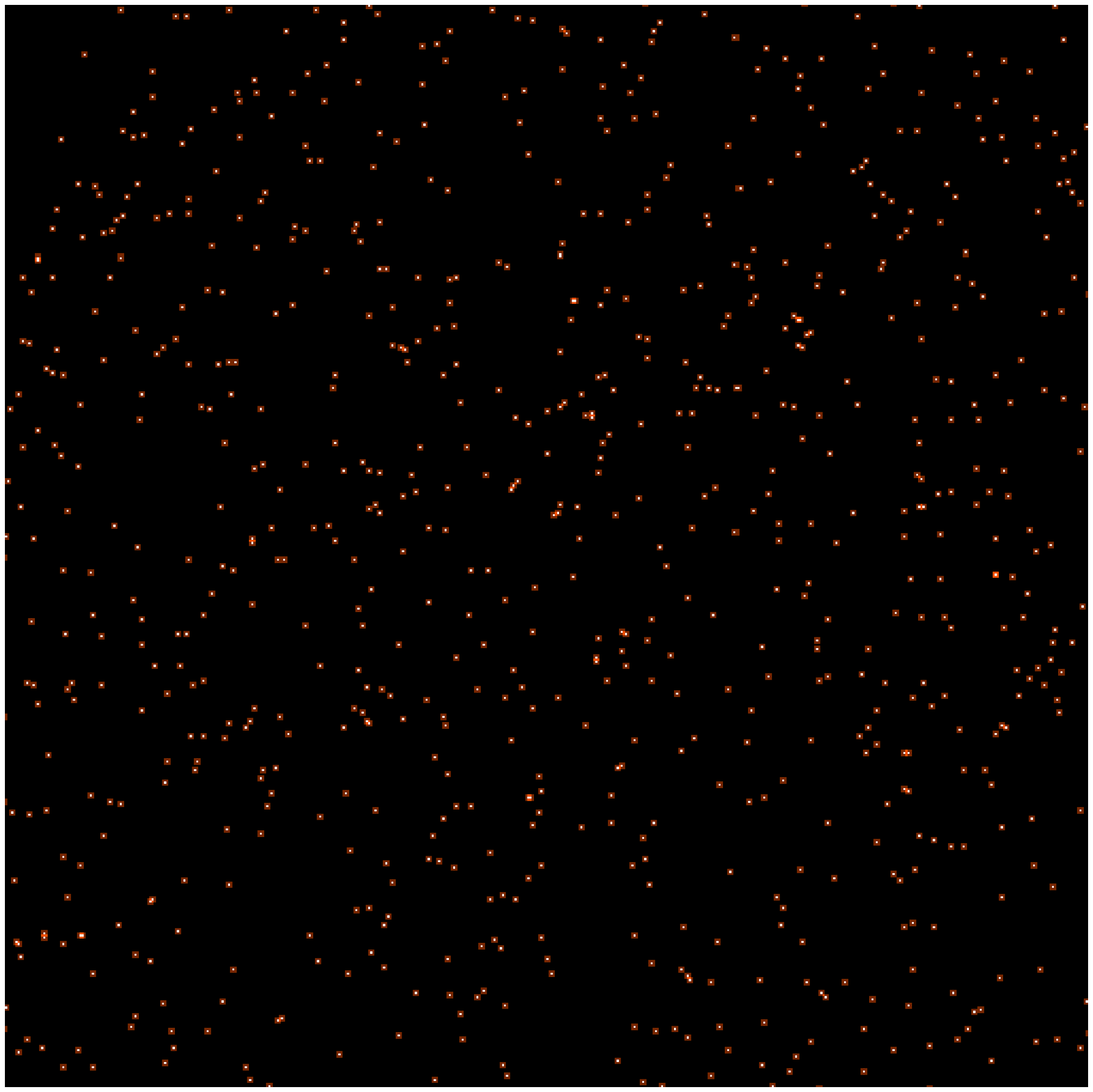}
\end{center}
\caption{Same as Fig.\,\ref{kdennerl-B1_fig:v00}, but for $E=$1.5\,--\,10~keV.
No trace of Venus is visible.}
\label{kdennerl-B1_fig:v583gr_}
\end{figure}

\begin{figure*}[!ht]
\begin{center}
\includegraphics[clip,width=17.8cm]{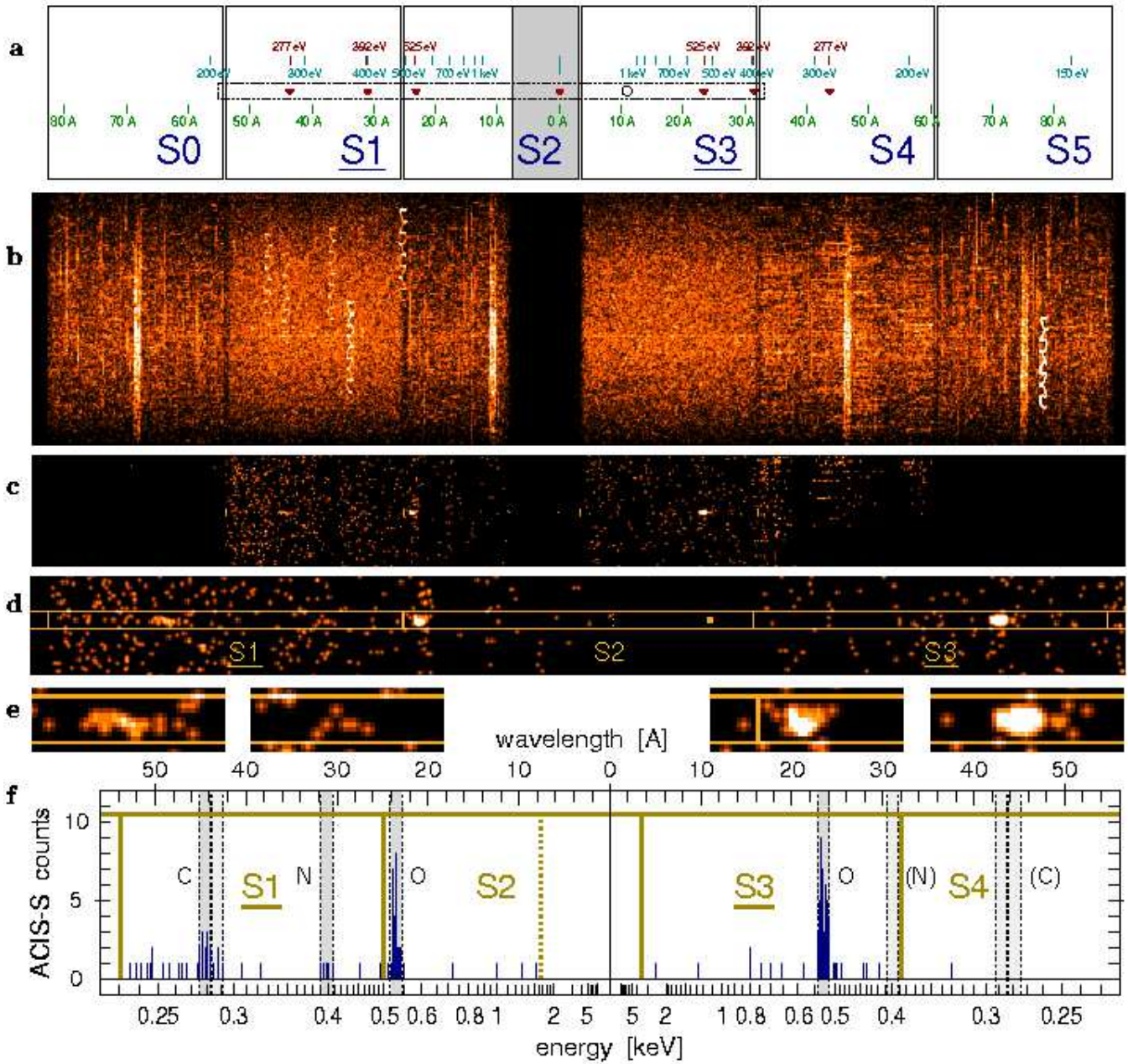}
\end{center}
\caption{
{\bf a)} Expected LETG spectrum of Venus on the ACIS--S array. S1 and S3 are
back--illuminated CCDs with increased sensitivity at low energies
(underlined), while the others are front--illuminated. The nominal aimpoint,
in S3, is marked with a circle. The aimpoint was shifted by $3\farcm25$ into
S2, to get more of the fluorescence lines covered by back--illuminated CCDs.
Energy and wavelength scales are given along the dispersion direction. During
the two pointings, Venus was moving perpendicularly to the dispersion
direction. In order to avoid saturation of telemetry due to optical light, the
shaded area around the zero order image in S2 was not transmitted. Images of
Venus are drawn at the position of the C, N, and O fluorescence lines, with the
correct size and orientation. The dashed rectangle indicates the section of
the observed spectrum shown below.
{\bf b)} Distribution of all events with standard grades from the first
ACIS--S pointing (obsid 2411), remapped into a coordinate system comoving with
Venus, with the Sun at bottom. The events were binned into $2\farcs5$
pixels. The white vertical curves are the trails of bright pixels, caused by a
superposition of the Lissajous pattern of the satellite attitude and the
apparent motion of Venus. The faint signal from Venus is lost in the noise.
{\bf c)} Central section of (b), after excluding bright pixels, and using the
intrinsic resolution of ACIS for selecting only photons in the appropriate
energy range. The two bright crescents symmetric to the center are images
of Venus in
the line of the O--K$_{\alpha}$ fluorescence emission, while the elongated
enhancement at left is at the position of the C--K$_{\alpha}$ fluorescence
emission line. The position of the zero order image (not transmitted) is
indicated by a dot in S2.
{\bf d,\,e)} Enlargements of (c).
{\bf f)} Spectral scan along the region outlined in (d). Scales are given in
keV and \AA. The observed C, N, and O fluorescence emission lines are enclosed
by dashed lines; the width of these intervals matches the size of the Venus
crescent ($22\farcs8$).
}
\label{kdennerl-B1_fig:letg_sum}
\end{figure*}

At the time of the observation, Venus was moving across the sky with a proper
motion of $2\farcm6$/hour. As the CCDs were read out every 3.2~s, there was no
need for continuous tracking. The spacecraft was oriented such that Venus
would move parallel to one side of the CCDs and perpendicular to the
dispersion direction in the LETG observation. To keep Venus well inside the
$8\farcm3$ field of view (FOV) of ACIS--S perpendicular to the dispersion
direction, Chandra was repointed at the middle of the LETG\,/\,ACIS--S
observation (Fig.\,\ref{kdennerl-B1_fig:acis-s},
Tab.\,\ref{kdennerl-B1_tab:obscxo}). For ACIS--I with its larger
$16\farcm9\times16\farcm9$ FOV, no repointing during the observation was
necessary (Fig.\,\ref{kdennerl-B1_fig:acis-i},
Tab.\,\ref{kdennerl-B1_tab:obscxo}).
As the photons were recorded time--tagged, an individual post--facto
transformation into the rest frame of Venus was possible. This is illustrated
in Figs.\,\ref{kdennerl-B1_fig:acis_i_trail_g} and \ref{kdennerl-B1_fig:v00}.
The fact that all observations were performed
with CCDs with intrinsic energy resolution made it possible to suppress
the background with high efficiency
(cf.\,Fig.\,\ref{kdennerl-B1_fig:v00}\,$\leftrightarrow\,%
$\ref{kdennerl-B1_fig:v583gr_} and 
Fig.\,\ref{kdennerl-B1_fig:letg_sum}, b\,$\leftrightarrow$\,c).

\clearpage

\begin{figure*}[!ht]
\begin{center}
\includegraphics[clip,width=17cm]{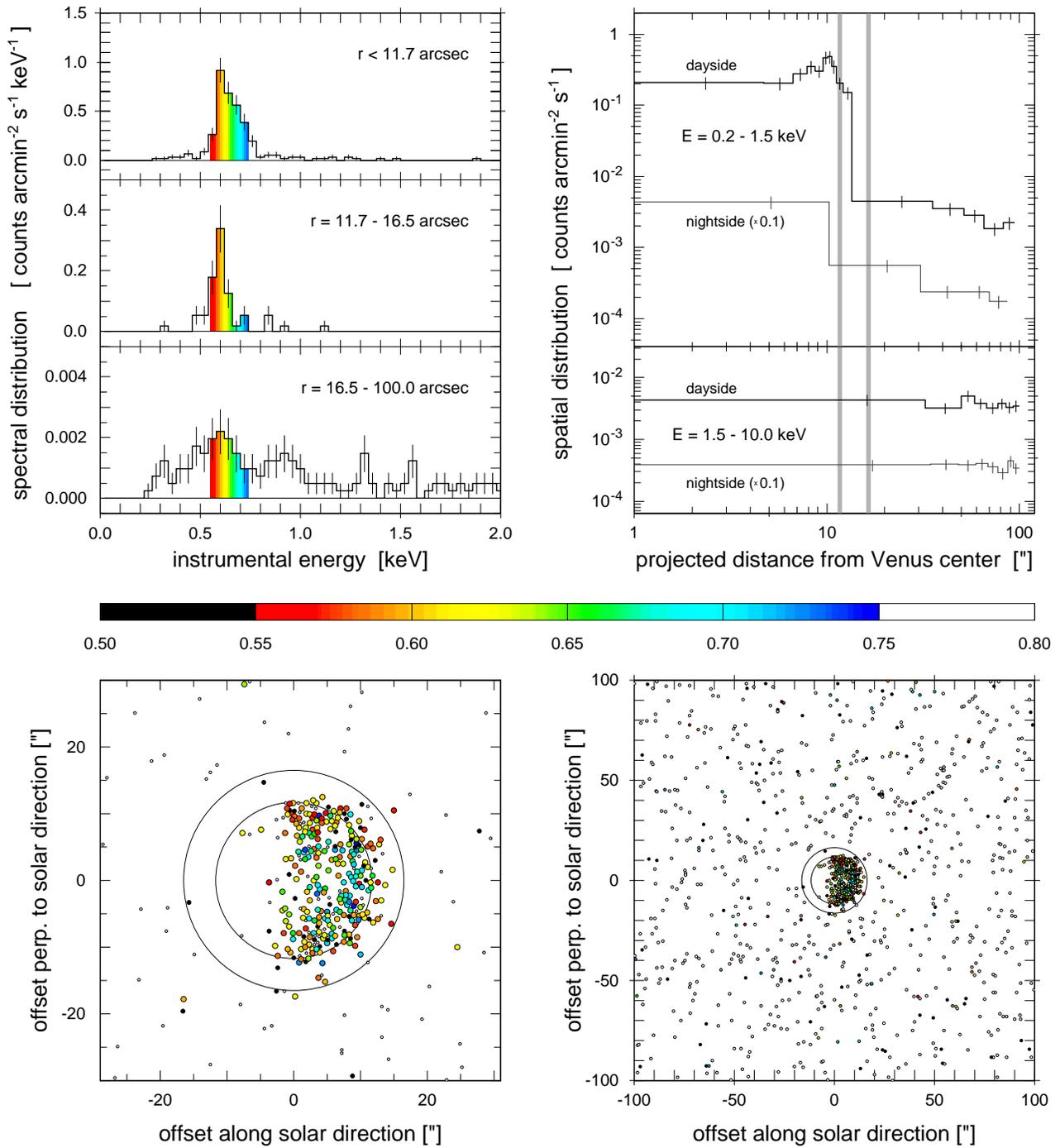}
\end{center}
\caption{Summary of the ACIS--I observation. The frames at bottom
show the distribution of photons around Venus in two different scales.
Photons in the instrumental energy range 0.55\,--\,0.75~keV are
highlighted by larger dots, filled with the color indicated above,
while photons of 0.20\,--\,0.55~keV and 0.75\,--\,15.7~keV are plotted
as black and white dots, respectively. In some cases the dots have
been slightly shifted (by typically less than $1''$) to minimize overlaps.
Two large circles are shown, the inner one, with $r=11\farcs7$,
indicating the geometric size of Venus, and the outer one, with
$r=16\farcs5$, enclosing practically all photons detected from Venus.
The frames at upper left show the energy spectra observed for
the three areas. The spectra of the two inner regions peak around
0.6~keV, with a tendency towards higher/lower energies in the inner/outer
region. This behaviour is most likely caused by optical loading, a
superposition of 0.53~keV photons and optical photons. The spectrum of the
outermost region, i.e., the area of $200''\times200''$ outside the outer
($r=16\farcs5$) circle, shows no evidence for line emission. The histograms at
upper right show the spatial distribution of the photons in the soft
($E=0.2$\,--\,1.5~keV) and hard ($E=1.5$\,--\,10.0~keV) energy range, in terms
of surface brightness along radial rings around Venus, separately for the
`dayside' (offset along solar direction $>0$) and the `nightside' (offset
$<0$). For better clarity the nightside histograms were shifted by one decade
downward. The bin size was adaptively determined so that each bin contains at
least 24 counts. Thick vertical lines mark the radii of $11\farcs7$ and
$16\farcs5$.}
\label{kdennerl-B1_fig:vn4spa_v2}
\end{figure*}

\clearpage

\section{Results}

The ACIS--I data clearly show that the X--ray spectrum of Venus is very soft:
at energies $E>1.5\mbox{ keV}$ no enhancement is seen at the position of
Venus, neither in the image (Fig.\,\ref{kdennerl-B1_fig:v583gr_}) nor in the
surface brightness profile (Fig.\,\ref{kdennerl-B1_fig:vn4spa_v2}). We
determine a $3\,\sigma$ upper limit of $2.5\cdot10^{-4}\mbox{ counts/s}$ to
any flux from Venus in the 1.5\,--\,2.0~keV energy range. The corresponding
value for $E=2$\,--\,8~keV is $5.6\cdot10^{-4}\mbox{ counts/s}$. Further
spectral analysis of the ACIS--I data is complicated by the presence of
optical loading (Fig.\,\ref{kdennerl-B1_fig:vn4spa_v2}).

The LETG spectrum, however, which is completely uncontaminated by optical
light, clearly shows that most of the observed flux comes from O--K$_\alpha$
fluorescence (Fig.\,\ref{kdennerl-B1_fig:letg_sum}).
As this flux is monochromatic, images of the Venus crescent
(illuminated from bottom) show up along the dispersion direction.
In addition to the O--K$_\alpha$
emission, we detect also fluorescence emission from C--K$_\alpha$ and,
marginally, from N--K$_\alpha$. The C--K$_\alpha$ image appears elongated, and
spectral fits indicate the presence of an additional emission line at
0.29~keV, which might be the signature of the C $1s\to\pi^{\star}$ transition
in CO$_2$ and CO.

Compared to its optical flux $f_{\rm opt}$, the total X--ray flux $f_{\rm x}$
from Venus is extremely low: $f_{\rm x}= 2\cdot10^{-10}\,f_{\rm opt}$.
Taking into account that the energy of a K$_{\alpha}$ photon exceeds
that of an optical photon by two orders of magnitude, we find that on
average there is only one X--ray photon among $5\cdot10^{11}$ photons
from Venus. This extremely small fraction of X--ray versus optical
flux, combined with the soft X--ray spectrum and the proximity of
Venus to the Sun, illustrates the challenge of observing Venus in
X--rays. The X--ray flux is emitted in just three narrow emission
lines. Outside these lines the $f_{\rm x}/f_{\rm opt}$ ratio is even
orders of magnitude lower.

\begin{figure}[!hb]
\vspace*{-6mm}
\begin{center}
\includegraphics[clip,width=7.7cm]{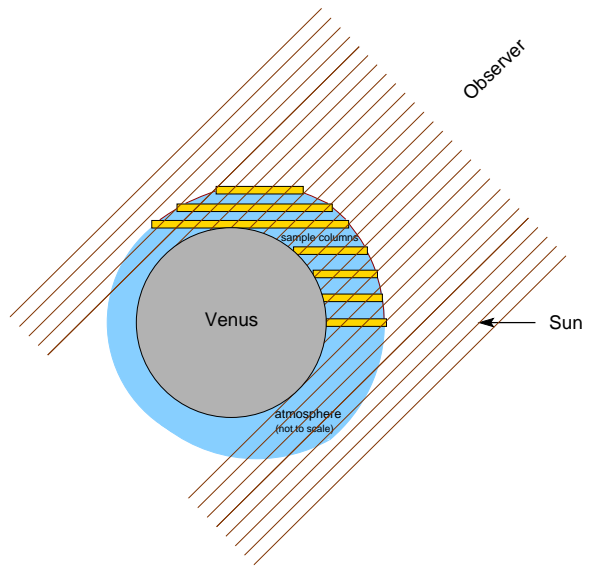}
\end{center}
\vspace*{-2mm}
\renewcommand{\baselinestretch}{0.9}
\caption{Modeling the X--ray appearance of Venus due to fluorescent
scattering of solar X--rays. In the first step, the incident solar spectrum is
computed for atmospheric columns which are parallel to the solar direction.
The calculation starts at the top of the column and propagates towards the
surface, taking the increasing attenuation into account. The absorbed flux is
then converted into volume emissivities due to C, N, and O fluorescence. By
sampling the emissivities in the volume elements along the line of sight,
starting from the volume element which is farthest away from the observer, and
taking the attenuation of this radiation due to subsequent photoabsorption
along the line of sight into account, images of Venus can be obtained in the
three fluorescence energies.
}
\label{kdennerl-B1_fig:vsim}
\end{figure}

\begin{figure*}[!ht]
\begin{center}
\includegraphics[clip,width=18cm]{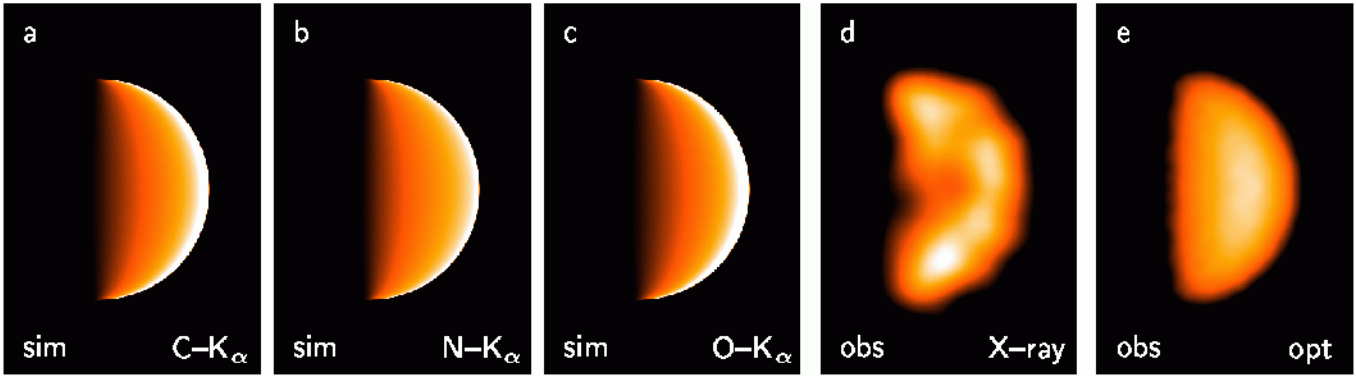}
\end{center}
\caption{{\bf a\,--\,c)} Simulated X--ray images of Venus at
C--K$_{\alpha}$, N--K$_{\alpha}$, and O--K$_{\alpha}$, for
a phase angle of $86.5^{\circ}$.
The X--ray flux is coded in a linear scale,
extending from zero (black) to
$1.2\cdot10^6\mbox{ ph cm}^{-2}\mbox{ s}^{-1}$ (a),
$5.2\cdot10^4\mbox{ ph cm}^{-2}\mbox{ s}^{-1}$ (b), and
$1.6\cdot10^6\mbox{ ph cm}^{-2}\mbox{ s}^{-1}$ (c),
(white). All images show considerable limb brightening,
especially at C--K$_{\alpha}$ and O--K$_{\alpha}$.
{\bf d)} Observed X--ray image: same as Fig.\,\ref{kdennerl-B1_fig:xfirst},
but smoothed with a Gaussian filter with $\sigma=1\farcs8$ and
displayed in the same scale as the simulated images. This image
is dominated by O--K$_{\alpha}$ fluorescence photons.
{\bf e)}~Optical image of Venus, taken by one of the authors (KD)
with a $4''$ Newton reflector on 2001 Jan 12.72 UT, 20 hours before
the ACIS--I observation (cf.\,Tab.\,\ref{kdennerl-B1_tab:obscxo}).}
\label{kdennerl-B1_fig:simsum}
\end{figure*}

In the X--ray image (Fig.\,\ref{kdennerl-B1_fig:v00}) the crescent of Venus is
clearly resolved and allows detailed comparisons with the optical appearance.
An optical image (Fig.\,\ref{kdennerl-B1_fig:simsum}\,e), taken at the same
phase angle, shows a sphere which is slightly more than half illuminated,
closely resembling the geometric illumination
(Fig.\,\ref{kdennerl-B1_fig:v7gg}), with the brightness maximum well inside
the crescent. In the X--ray image the sphere appears to be less than half
illuminated. The most striking difference, however, is the pronounced limb
brightening, which is particularly obvious in the surface brightness profiles
(Fig.\,\ref{kdennerl-B1_fig:vn4spa_v2}) and in the smoothed X--ray image
(Fig.\,\ref{kdennerl-B1_fig:simsum}\,d).

\begin{figure}[!ht]
\begin{center}
\unitlength=1cm
\begin{picture}(7.5,10)
\put(0,6.1){\includegraphics[clip,width=7.5cm]{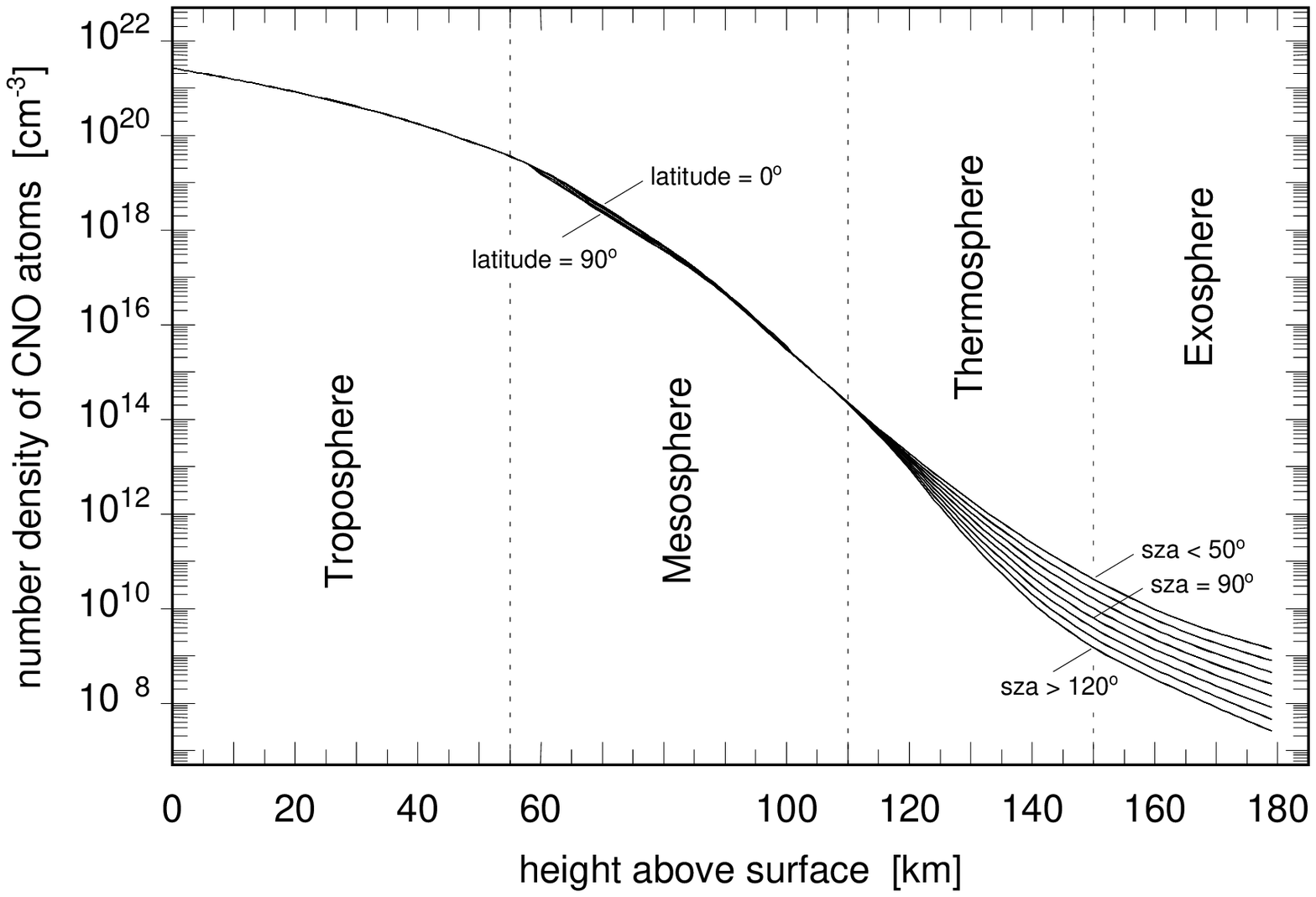}}
\put(0,0.1){\includegraphics[clip,width=7.44cm]{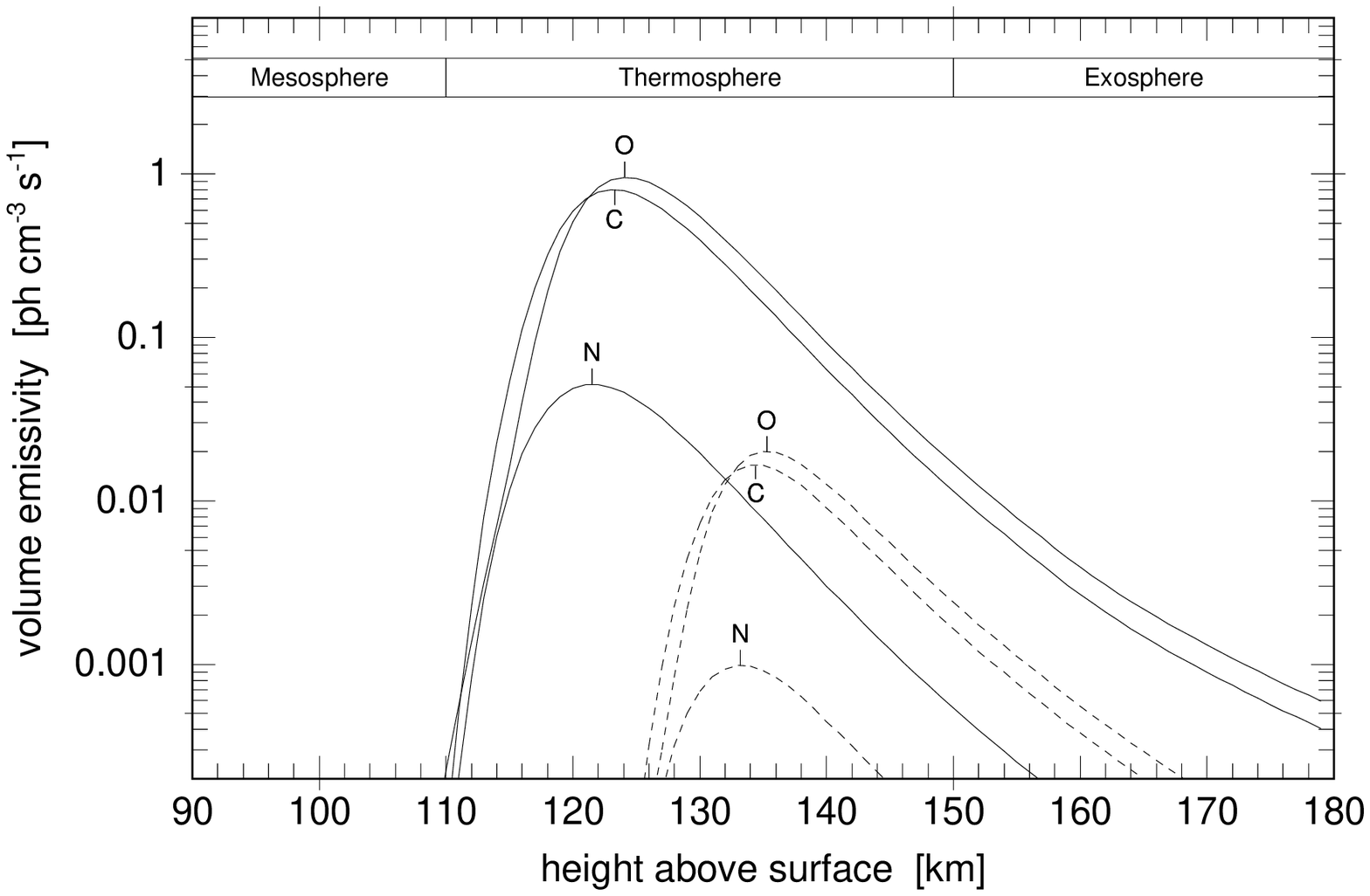}}
\put(3.0,6.15){\vector(-3,-2){1.75}}
\put(4.15,6.9){\line(-3,-2){0.6}}
\put(7.25,6.47){\vector(0,-1){1.47}}
\put(0.3,6.5){\large\bf a}
\put(0.3,4.9){\large\bf b}
\put(3.2,3.35){\tiny\sf subsolar}
\put(3.9,1.8){\tiny\sf terminator}
\end{picture}
\end{center}
\vspace*{-2mm}
\renewcommand{\baselinestretch}{0.9}
\caption{{\bf a)} Number density $n_{\rm CNO}=n_{\rm C}+n_{\rm N}+n_{\rm O}$
of the sum of C, N, and O atoms in the Venus model atmosphere as a
function of the height above the surface. Between 60 and 100~km, the
density shows a slight dependence on latitude. Above 110~km, the
density depends considerably on the solar zenith angle (sza).
{\bf b)} Volume emissivities of C, N, and O K$_{\alpha}$ fluorescence photons
at zenith angles of zero (subsolar, solid lines) and $90\degr$
(terminator, dashed lines) for the incident solar spectrum of
Fig.\,\ref{kdennerl-B1_fig:incflx}. The height of maximum emissivity rises with
increasing solar zenith angles because of increased path length and absorption
along oblique solar incidence angles. In all cases maximum emissivity occurs
in the thermosphere, where the optical depth depends also on the solar zenith
angle.}
\label{kdennerl-B1_fig:atmo1}
\end{figure}

\begin{figure}[!ht]
\begin{center}
\includegraphics[clip,width=8.5cm]{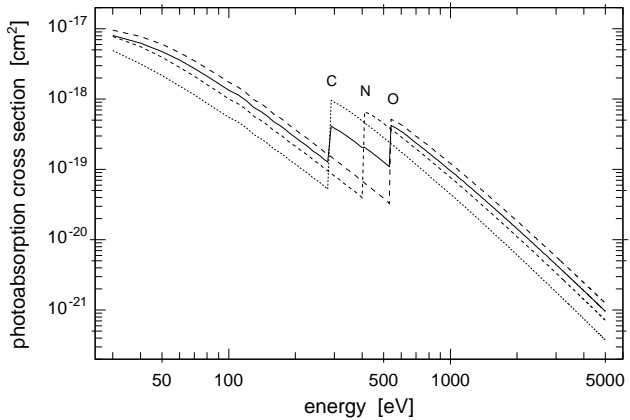}
\end{center}
\vspace*{-3mm}
\caption{Photoabsorption cross sections $\sigma_{\rm C}$,
$\sigma_{\rm N}$, $\sigma_{\rm O}$ for C, N, and O (dashed lines),
and $\sigma_{\rm CNO}$ for the chemical composition of the Venus
atmosphere (solid line).}
\label{kdennerl-B1_fig:crsc}
\end{figure}

\begin{figure}[!ht]
\vspace*{-10mm}
\begin{center}
\includegraphics[clip,height=9cm,width=6.5cm,angle=-90]{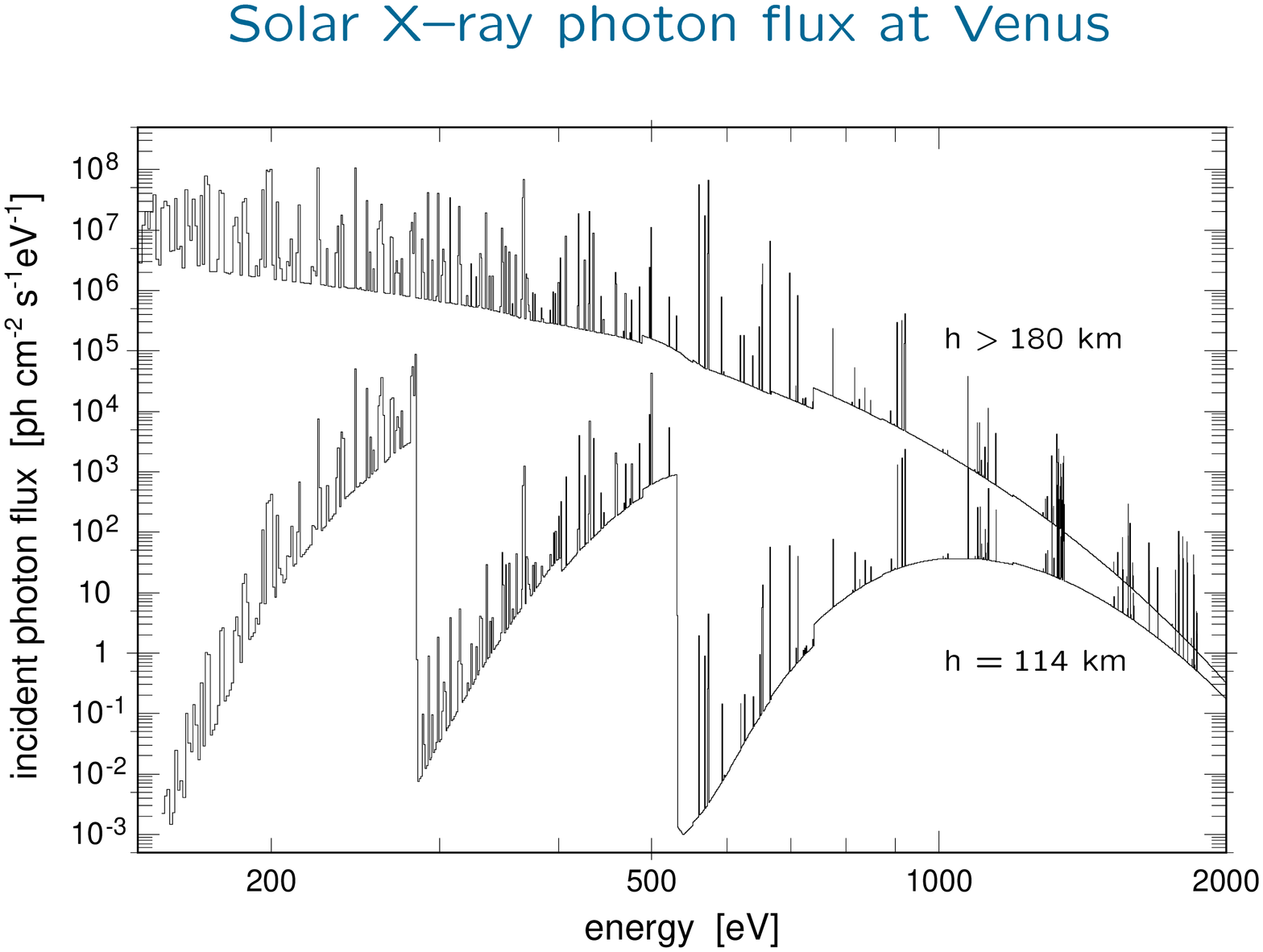}
\end{center}
\vspace*{-2mm}
\renewcommand{\baselinestretch}{0.9}
\caption{Incident solar X--ray photon flux on top of the Venus atmosphere
($h>180\mbox{ km}$) and at 114~km height (along subsolar direction). The
spectrum is plotted with a bin size of 1~eV, which we used for the
simulation in order to preserve the spectral details. At 114~km, it is
considerably attenuated just above the K absorption edges, recovering towards
higher energies.}
\vspace*{10mm}
\label{kdennerl-B1_fig:incflx}
\end{figure}

\begin{figure}[!ht]
\vspace*{-8mm}
\begin{center}
\includegraphics[clip,width=8.5cm,height=6.5cm]{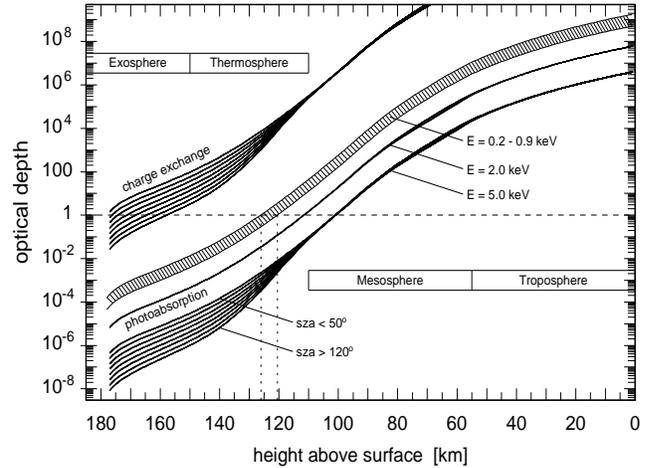}
\end{center}
\vspace*{-2mm}
\renewcommand{\baselinestretch}{0.9}
\caption{Optical depth $\tau_{\rm CNO}=\tau_{\rm C}+\tau_{\rm N}+\tau_{\rm O}$
of the Venus model atmosphere with respect to CNO photoabsorption, as seen
from outside. The upper/lower boundaries of the hatched area refer to energies
just above/below the C and O edges (cf.\ Fig.\,\ref{kdennerl-B1_fig:crsc}).
For better clarity the dependence on the solar zenith angle (sza) is only
shown for $E=5.0\mbox{ keV}$; the curves for the other energies refer to
$\mbox{sza}<50^{\circ}$. The dashed line, at $\tau=1$, marks the transition
between the transparent ($\tau<1$) and opaque ($\tau>1$) range. For a specific
energy, the optical depth increases by at least 13 orders of magnitude between
180~km and the surface. For comparison, the collisional depth resulting from
charge exchange interactions with highly ionized atoms in the solar wind
is also shown, for which a constant cross section of
$3\cdot10^{-15}\mbox{ cm}^2$ was assumed. Due to this large cross section, the
tenuous exosphere of Venus is already collisionally thick. The flux of highly
charged heavy solar wind ions, however, is too small to contribute
significantly to the X--ray flux from Venus.}
\vspace*{5mm}
\label{kdennerl-B1_fig:atmo2}
\end{figure}

\begin{figure*}[!ht]
\begin{center}
\includegraphics[clip,width=15cm]{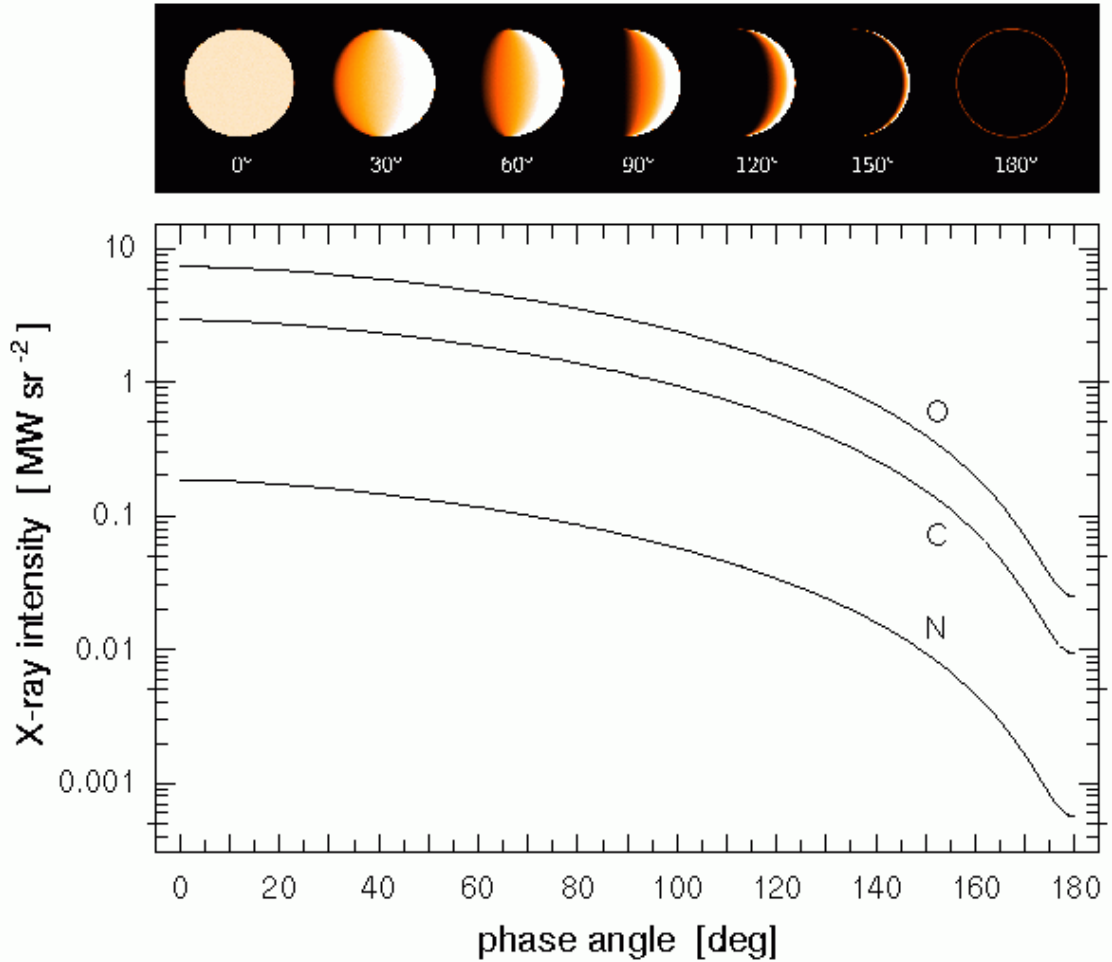}
\end{center}
\caption{X--ray intensity of Venus as a function of phase angle, in the
fluorescence lines of C, N, and O. The images at top, all displayed in the
same intensity coding, illustrate the appearance of Venus at O--K$_{\alpha}$
for selected phase angles.}
\label{kdennerl-B1_fig:vnlum1sim_1}
\end{figure*}

For a quantitative understanding of this brightening we performed a numerical
simulation of the appearance of Venus in soft X--rays, based on fluorescent
scattering of solar X--ray radiation (Fig.\,\ref{kdennerl-B1_fig:vsim}). The
ingredients to this simulation were the composition and density structure of
the Venus atmosphere (Fig.\,\ref{kdennerl-B1_fig:atmo1}a), the photoabsorption
cross sections (Fig.\,\ref{kdennerl-B1_fig:crsc}) and fluorescence
efficiencies of the major atmospheric constituents, and the incident solar
spectrum (Fig.\,\ref{kdennerl-B1_fig:incflx}). The simulation showed that the
volume emissivity peaks at heights of 120\,--\,140~km and extends into the
tenuous, optically thin parts of the thermosphere and exosphere
(Fig.\,\ref{kdennerl-B1_fig:atmo1}b). We see the volume emissivities
accumulated along the line of sight without considerable absorption
(Fig.\,\ref{kdennerl-B1_fig:atmo2}), so that
the observed brightness is mainly determined by the extent of the atmospheric
column along the line of sight. This causes the pronounced brightening at the
sunward limb, accompanied by reduced brightness at the terminator. Limb
brightening would also be observed at other phase angles
(Fig.\,\ref{kdennerl-B1_fig:vnlum1sim_1}).

The simulation shows that the amount of limb brightening is different for
the three fluorescence energies (Fig.\,\ref{kdennerl-B1_fig:simsum} a\,--\,c)
and depends on the properties of the Venus atmosphere at heights above 110~km.
Thus, information about the chemical composition and density structure of the
Venus thermosphere and exosphere can be obtained by measuring the X--ray
brightness distribution across the planet in the individual K$_{\alpha}$
fluorescence lines. This opens the possibility of using X--ray observations
for remotely monitoring the properties of these regions in the Venus atmosphere
which are difficult to investigate otherwise, and their response to solar
activity.

\clearpage

\begin{thebibliography}{}

\bibitem[\protect\astroncite{Dennerl}{2002}]{kdennerl-B1:den02}
Dennerl, K., Burwitz, V., Englhauser, J., Lisse, C., Wolk, S. 2002,
A\&A 386, 319\,--\,330

\end{thebibliography}
\end{document}